\documentclass{PoS}
\usepackage{multirow}
\def\beq{\begin{equation}}
\def\eeq{\end{equation}}
\newcommand{\ssi}{\sigma^{\rm SI}_p}
\newcommand{\gev}{\,\, \mathrm{GeV}}
\newcommand{\cha}[1]{\tilde \chi^\pm_{#1}}
\newcommand{\staue}{\tilde \tau_1}
\newcommand{\tb}{\tan\beta}
\newcommand{\ETslash}{\ensuremath{/ \hspace{-.7em} E_T}}
\newcommand{\model}[3]{${\tt #1}_{\tt\bf #2}^{\tt #3}$}
\newcommand{\DM}[2]{${\tt #1}_{\tt #2}$}
\newcommand{\bhline}[1]{\noalign{\hrule height #1}}

\title{Supersymmetric Dark Matter or Not}

\ShortTitle{Supersymmetric Dark Matter or Not}

\author{\speaker{Keith A. Olive}\\
       William I. Fine Theoretical Physics Institute, School of Physics and Astronomy,\\
University of Minnesota, Minneapolis, MN 55455,\,USA\\
        E-mail: \email{olive@umn.edu}}

\abstract{ The lack of evidence for low energy supersymmetry at the LHC implies a 
supersymmetry scale in excess a TeV. While this is consistent (and even helpful) with
a Higgs boson mass at $\approx$ 125 GeV, simple supersymmetric models with scalar
and gaugino mass universality are being pushed into strips of parameter space.
These often require coannihilations to obtain an acceptable relic density and the
extent of these coannihilation strips will be discussed. 
In contrast, non-supersymmetric grand unified theories such as SO(10) may
also provide a dark matter candidate.  Because of the presence of an intermediate scale,
these theories may unify gauge couplings, provide for neutrino masses and a suitably long lived proton.}

\FullConference{The 11th International Workshop Dark Side of the Universe 2015\\
		14-18 December 2015\\
		Kyoto, Japan}

\begin{document}

\section{Before Run I}

Despite its relative simplicity as an extension of the Standard Model (SM), 
there are many possible realizations of low energy supersymmetry
which can be traced to the unknown mechanism of supersymmetry breaking
and the multitude of parameters associated with that breaking. The most widely
studied version of low energy supersymmetry makes some strong assumptions
concerning these supersymmetry breaking parameters. Namely,
it assumes that all gaugino masses are universal at some input
renormalization scale (usually taken to be the Grand Unification (GUT) 
scale at which gauge coupling unification 
occurs). It further assumes that all supersymmetry breaking scalar masses and
trilinear terms are also universal at the same input scale. This constrained version of the
minimal supersymmetric standard model (CMSSM) \cite{funnel,cmssm,elos,eelnos} is a 4-parameter theory defined by the
gaugino mass, $m_{1/2}$, the scalar mass, $m_0$, the trilinear mass term, $A_0$,
and the ratio of the two Higgs vacuum expectation values, $\tan \beta$.
In the CMSSM, one uses the conditions derived
by the minimization of the Higgs potential after radiative electroweak symmetry breaking to solve for
the Higgs mixing mass, $\mu$ and the bilinear mass term $B_0$ (or equivalently $\mu$ and the Higgs pseudoscalar mass, $m_A$) for fixed $\tan \beta$.

As is now well known, Run I of the LHC did not provide evidence for supersymmetry,
and instead pushed supersymmetric mass scales to higher energies of order a TeV.
At the higher mass scales, viable dark matter models rely on enhanced 
annihilations in order to obtain the correct relic density.
This may due to either coannihlations \cite{gs}, direct s-channel annihilations on a pole (the funnel
region) \cite{gs}
or in the focus point where the $\mu$ parameter is relatively small \cite{fp}. 
The coannihilation and funnel mechanisms require very special choices of the input parameters
which cause near degeneracies among the sparticle masses. Thus regions of good relic
density typically lie in strips in a ($m_{0}, m_{1/2}$) plane (the focus point also forms a strip along the
boundary where radiative symmetry breaking fails). The ultimate potential for the discovery of supersymmetry
depends on the extent of these strips.

An alternative to supersymmetric dark matter is provided by SO(10)
grand unification \cite{so10,GN2}. The presence of an intermediate scale
allows for the possibility of gauge coupling unification (without supersymmetry) \cite{GN2,so10gc,moqz,mnoqz,noz,mnoz}
and the breaking of the intermediate scale provides a natural origin for the see-saw mechanism 
for generating neutrino masses \cite{seesaw}. If the intermediate scale is broken
by a {\bf 126}-dimensional Higgs representation, the theory preserves a $Z_2$ symmetry \cite{Kibble:1982ae}
which can account for the stability of a new dark matter candidate \cite{moqz,mnoqz,noz,mnoz,so10dm,enoz}. 

\begin{figure}
  \includegraphics[height=.35\textwidth]{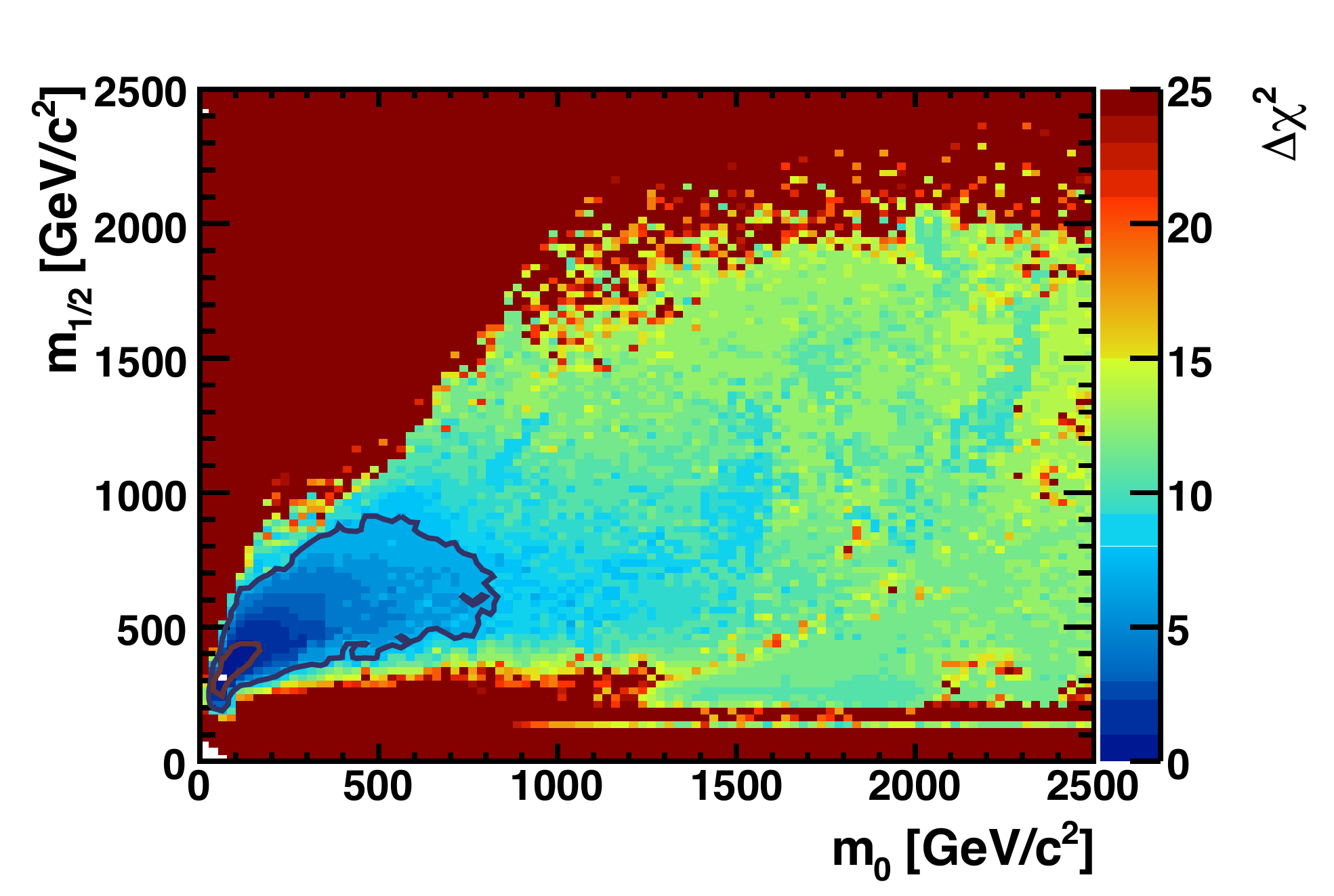}
   \includegraphics[height=.35\textwidth]{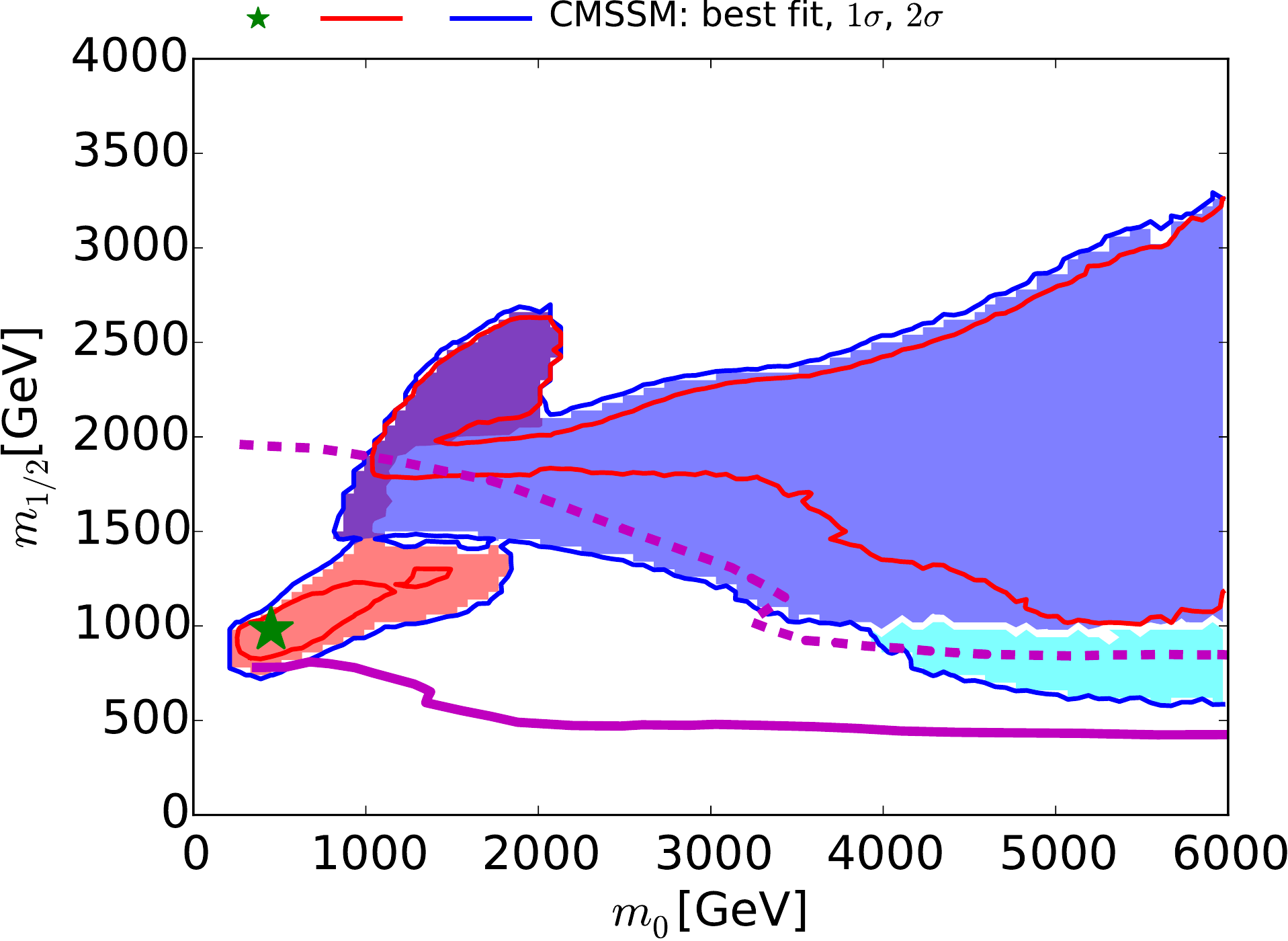}
  \caption{The $\Delta\chi^2$ functions in the $(m_0, m_{1/2})$ planes for
  the CMSSM from a mastercode frequentist analysis. The pre-LHC result is shown in the 
  left panel \cite{mc3}. Red and blue contours correspond to 68\% and 95\% CL contours and the best fit point is 
  depicted by a white dot. The post-LHC result is shown in the right panel \cite{mc12} using 8 TeV data at 20 fb$^{-1}$. Here the best fit point is shown by the filled star. The color of the shaded region
  indicates the dominant annihilation mechanism for obtained the correct relic density:
  stau coannihilation-pink; $A/H$ funnel-blue; focus point-cyan; and a hybrid region of stau coannihilation and funnel-purple. The solid and dashed purple curves show the run I reach and the expected run II reach at
  14 TeV at 3000 fb$^{-1}$ respectively. The latter corresponds approximately to the 95\% CL exclusion sensitivity with 300/fb at 14 TeV. } 
  \label{mcm0m12}
\end{figure}

Prior to Run I of the LHC, there was considerable excitement about the prospect for discovering 
supersymmetry as supersymmetric models such as the CMSSM
provided definite improvements to low energy precision phenomenology.
This can be seen in the left panel of Fig. \ref{mcm0m12} which shows the 
results of mastercode \cite{mc3,mc} - a 
frequentist Markov Chain Monte Carlo analysis of low energy experimental observables in
the context of supersymmetry. The figure shows the color coded values of $\Delta \chi^2$
relative to the best fit point shown by the white dot at low $m_{1/2}$ and low $m_0$.
Marginalization over $A_0$ and $\tan \beta$ was performed to produce this $(m_0, m_{1/2})$ plane.
The best-fit CMSSM point lies at
$m_0 = 60 \gev$,  $m_{1/2} = 310 \gev$,  $A_0 = 130 \gev$, $\tb = 11$
yielding the overall $\chi^2/{\rm N_{\rm dof}} = 20.6/19$ (36\% probability) 
with $m_h = 114.2$ GeV. Recall that this was a pre-LHC prediction and uses no LHC data.
Rather, it is based on a wide array of low energy observables including $(g_\mu -2)$, $M_W$, 
$B \to \tau \nu$, $b \to s \gamma$, the LEP limit on the Higgs mass, forward-backward asymmetries
among others (for a full list of observables used see \cite{mc3}). The relatively low value of $m_h$
was a common prediction of MSSM models \cite{Ellis:1990nz}. Indeed a dedicated scan for the distribution 
of Higgs masses in the CMSSM was made in \cite{ENOS} which found that when 
all phenomenological constraints (with or without $(g_\mu -2)$) are included, all models yielded
$m_h \le 128$ GeV. When $(g_\mu -2)$ is included, only models with $m_h < 126$ GeV were found.
Note that the scan sampled scalar and gaugino masses only out to 2 TeV.

\begin{figure}
  \includegraphics[height=.35\textwidth]{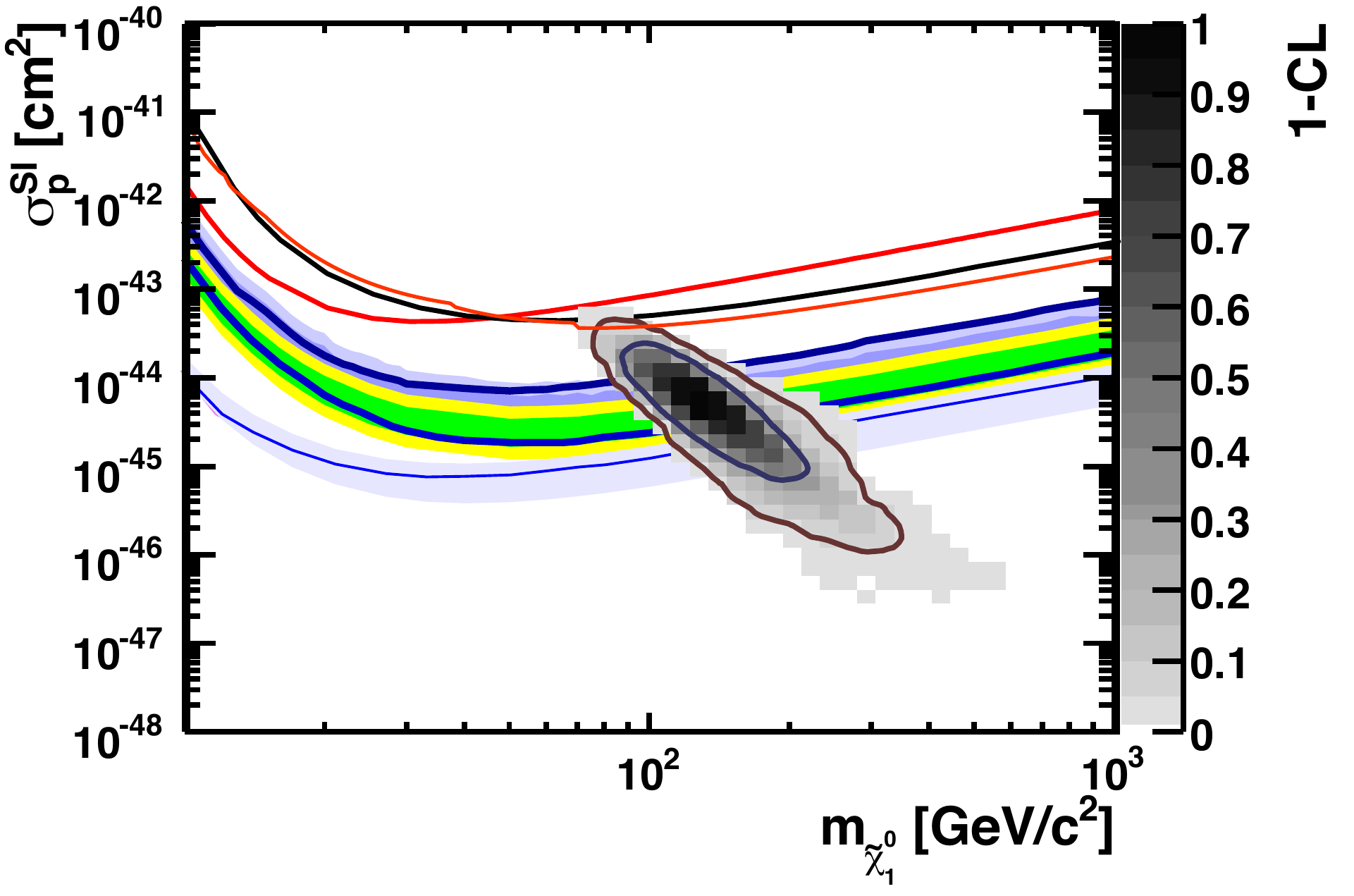}
   \includegraphics[height=.35\textwidth]{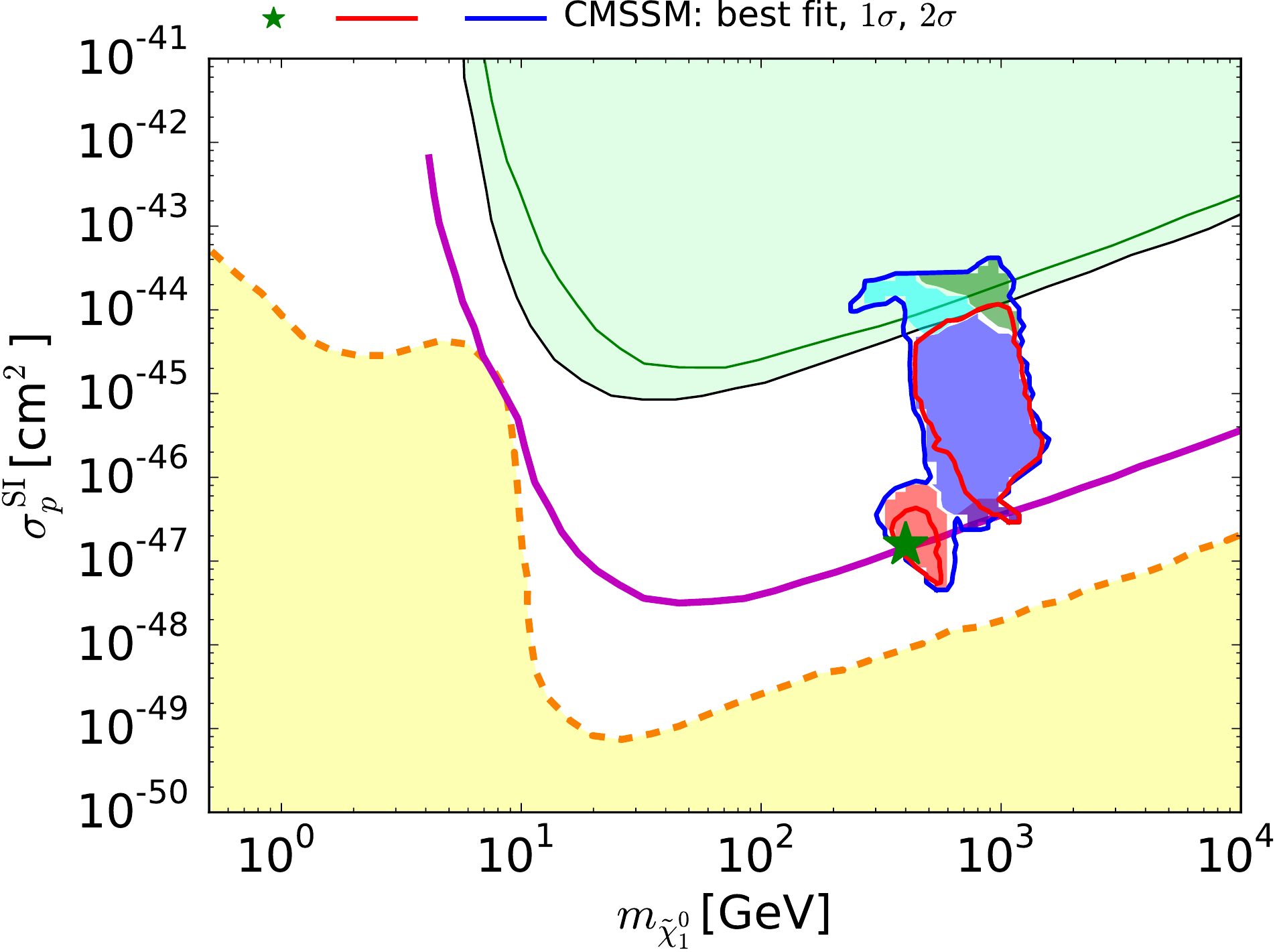}
  \caption{(left) The pre-LHC prediction for the spin-independent DM
scattering cross section, $\ssi$,  versus $m_\chi$
in the CMSSM \cite{mc3}. The solid lines
are the pre-LHC experimental upper limits from CDMS~\cite{CDMS} and XENON10\cite{Xe10},
while the bands are the more recent limits from XENON100 \cite{xe100100,xe100} and LUX \cite{lux}.   
(right) The post-run I likelihood contours for $\ssi$ \cite{mc12}.  Shading within the likelihood contours is the same as in
Fig. \protect\ref{mcm0m12}, though here we also see  a region where chargino coannihilations are dominant (green).
The green and black lines show the current sensitivities of the XENON100~\protect\cite{xe100}
and LUX~\protect\cite{lux} experiments,
respectively, and  the solid purple lines show the projected 95\% exclusion sensitivity of the LUX-Zepelin
(LZ) experiment~\protect\cite{LZ}.  The dashed orange line shows the
astrophysical neutrino `floor'~\protect\cite{nuback,cushman}, below which astrophysical neutrino backgrounds dominate (yellow region).
 }
  \label{mcssi}
\end{figure}

This optimism of discovering supersymmetry spread to the prospects of discovering dark matter
in direct detection experiments. The left panel of Fig.~\ref{mcssi} displays the 
pre-LHC preferred range of the
spin-independent DM scattering cross section $\ssi$
(calculated here assuming an optimistic $\pi$-N scattering term
$\Sigma_N = 64$~MeV) as a function of $m_\chi$ \cite{mc3}. 
We see that
the expected range of $\ssi$ lies just below the then present
experimental upper limits (solid lines) \cite{CDMS,Xe10}.
As one can see from the successive lower upper limits from later experiments \cite{xe100100,xe100,lux}
shown by the bands, these pre-LHC values for the elastic scattering cross section showed great 
promise for discovery.

\section{After Run I}

As Run I at the LHC, progressed, it was becoming increasingly clear that
supersymmetry if present at low energy at all, was at mass scales
larger than originally anticipated \cite{postlhc}.
Subsequent to the final run at the LHC, the picture looked very different.
In the right panel of Fig. \ref{mcm0m12}, the post-Run I likelihood contours in the $(m_0, m_{1/2})$ plane \cite{mc12}
are shown using 8 TeV results at
20 fb$^{-1}$ \cite{ATLAS20}. The best fit point based on the 8 TeV data is shown by the
filled star at (420,970) GeV with $A_0 = 3000$ GeV and $\tb = 14$, though the likelihood function
is quite flat and the exact position of the best point is not particularly meaningful. The $\chi^2/N_{\rm dof}$
is now  increased to 35.1/23 (5.1\% probability). This result may be compared with the Standard Model fit which yields
$\chi^2/N_{\rm dof}$
of 36.5/24 (5.0\% probability), which of course ignores the fact that there is no dark matter candidate in the Standard 
Model. 

We see in the right panel of Fig.~\ref{mcm0m12} that three DM mechanisms dominate in the CMSSM:
${\staue}$ coannihilation at low $m_0 \lesssim 2000 \gev$, the $H/A$ funnel at larger $m_0$ and $m_{1/2}$,
and the focus point at larger $m_0$ and smaller $m_{1/2}$ where the neutralino becomes 
a `well-tempered' mixture of bino and Higgsino~\cite{ADG}. There is also a hybrid ${\staue}/A/H$ region
extending up to $(m_0, m_{1/2}) \sim (2000, 2500) \gev$. 

As one can see from the right panel of Fig. \ref{mcssi}, there is still hope for direct detection experiments though the new best fit point implies a cross section of $\sim 10^{-47}$ cm$^2$, nearly two orders of magnitude below the current upper bound.  As commented on previously, the likelihood function
is rather flat between $10^{-47}$ cm$^2 \lesssim \ssi\ \lesssim 10^{-45}$~cm$^2$.
Note that in this case, we have adopted $\Sigma_{\pi N} = 50 \pm 
7$ MeV. In addition to the model results,
 the 90\% CL upper limits on $\ssi\ $ given by the XENON100 and LUX
 experiments~\cite{xe100,lux} as well as the expected reach from LZ \cite{LZ} are also displayed. The level of the atmospheric neutrino background~\cite{nuback,cushman} is shown by the shaded region at small cross sections.
The current XENON100 and LUX data already put strong pressure
on models where the focus-point or chargino coannihilation mechanism dominates. 
There are borderline regions that are formally
excluded by the $\ssi$ data considered in isolation, but become permitted at the 95\% CL in a global fit including other
observables, and also
{due to the uncertainties in the calculation of $\ssi$ that
have been included in the evaluation of the global $\chi^2$ function~\cite{MC9}}. 
We also see that the {$\cha1$ coannihilation region and most of the $H/A$ funnel region} would be accessible to
the planned LZ experiment. However, much of the ${\staue}$ coannihilation region lies below the LZ sensitivity,
though it could be accessible to a 20-tonne DM experiment such as
Darwin~\cite{Darwin}.

\section{The Strips}
As noted earlier, regions of parameter space with the correct relic density
typically occur in relatively narrow strips formed by special relations among the input parameters. 
In these cases, the relic density is strongly affected by coannihilations, or rapid $s$-channel annihilations.

\begin{figure}
  \includegraphics[height=.45\textwidth]{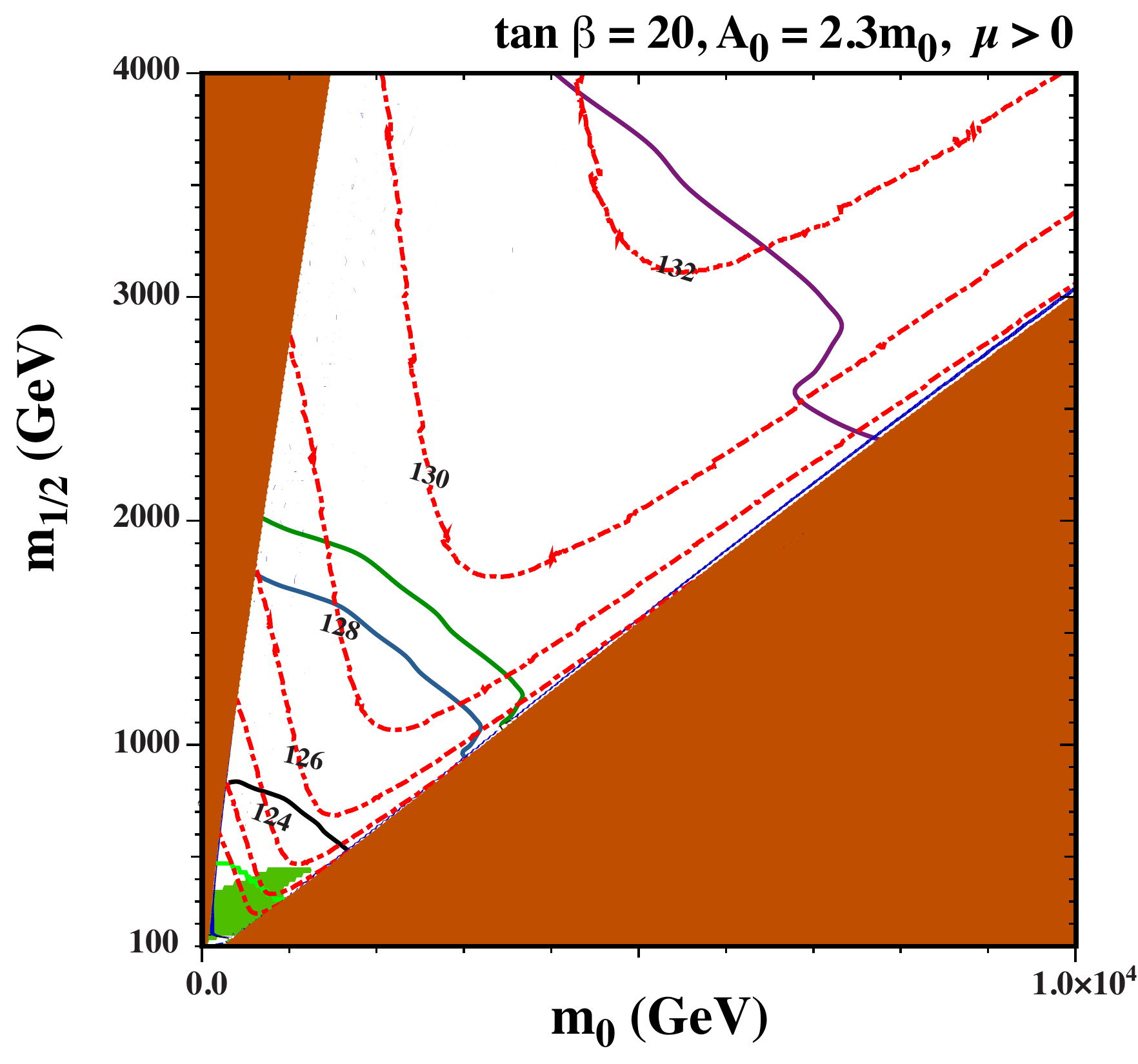}
   \includegraphics[height=.45\textwidth]{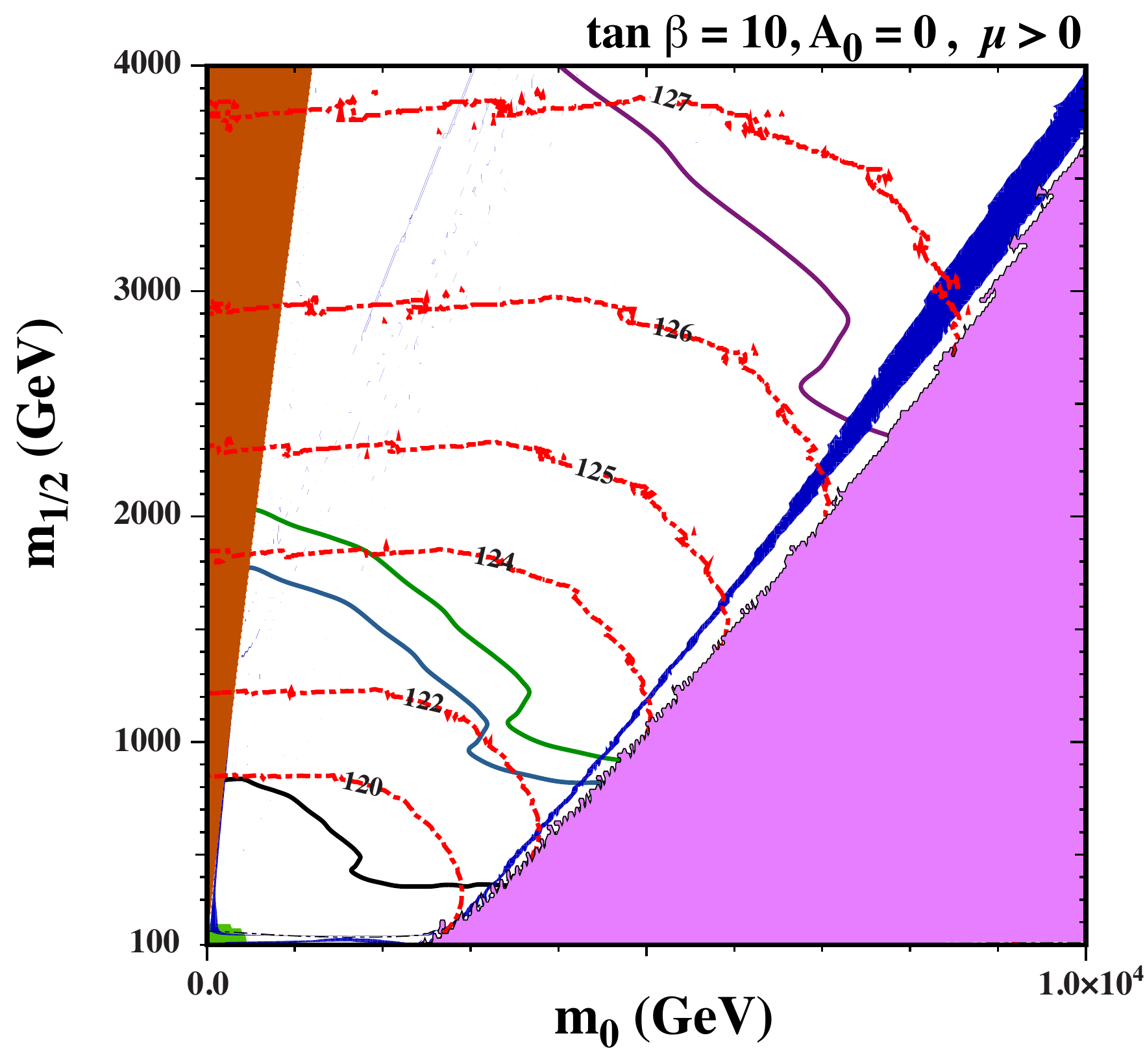}
  \caption{(left)  The $(m_0, m_{1/2})$ plane for fixed $\tan \beta = 20$ and $A_0 = 2.3 m_0$.
  (right)  The $(m_0, m_{1/2})$ plane for fixed $\tan \beta = 10$ and $A_0 = 0$.
  The dark red shaded regions are excluded because of a charged LSP and/or a tachyon,
 and the green region are excluded by $b \to s \gamma$ decay.
 There is no consistent electroweak vacuum
in the purple region in the right panel. 
 In the dark blue strips the relic LSP density lies within
the range allowed by cosmology, and the dashed red lines are contours of $m_h$ as calculated using
{\tt FeynHiggs~2.10.0} \cite{FH}. The solid black, blue, green and purple lines in each panel are
particle exclusion reaches for $\ETslash$ searches with the LHC at 8~TeV,
300 and 3000/fb with LHC at 14~TeV, and 3000/fb with HE-LHC at 33~TeV, respectively.
 }
  \label{mom12}
\end{figure}

The stau coannihilation strip \cite{stau-co,celmov}  is present when the mass of the lighter stau is nearly degenerate with
the lightest supersymmetric particle (LSP) which is often the bino in CMSSM models. 
An example (barely) showing the stau coannihilation strip is found in Fig. \ref{mom12}
which shows the $(m_0, m_{1/2})$ plane for fixed $\tan \beta = 20$ and $A_0 = 2.3 m_0$.
In the dark red shaded region at small $m_0$ extending to large $m_{1/2}$,
the lighter stau is the LSP and that region is excluded. Along the border of that
region, the stau and lightest neutralino are degenerate. The stau coannihilation strip
tracks that boundary up to roughly $m_{1/2} = 1$ TeV and is shown as a blue shaded strip.
Along the strip, the Higgs mass (shown by the red dot dashed curves computed with FeynHiggs \cite{FH}) does not exceed 124 GeV. 
The current and future reach of the LHC is shown by the solid black, blue, green and purple lines which are
particle exclusion reaches for $\ETslash$ searches with 20/fb at 8~TeV,
300 and 3000/fb at 14~TeV, and 3000/fb at a prospective HE-LHC at 33~TeV, respectively \cite{interplay}.

\begin{figure}
  \includegraphics[height=.45\textwidth]{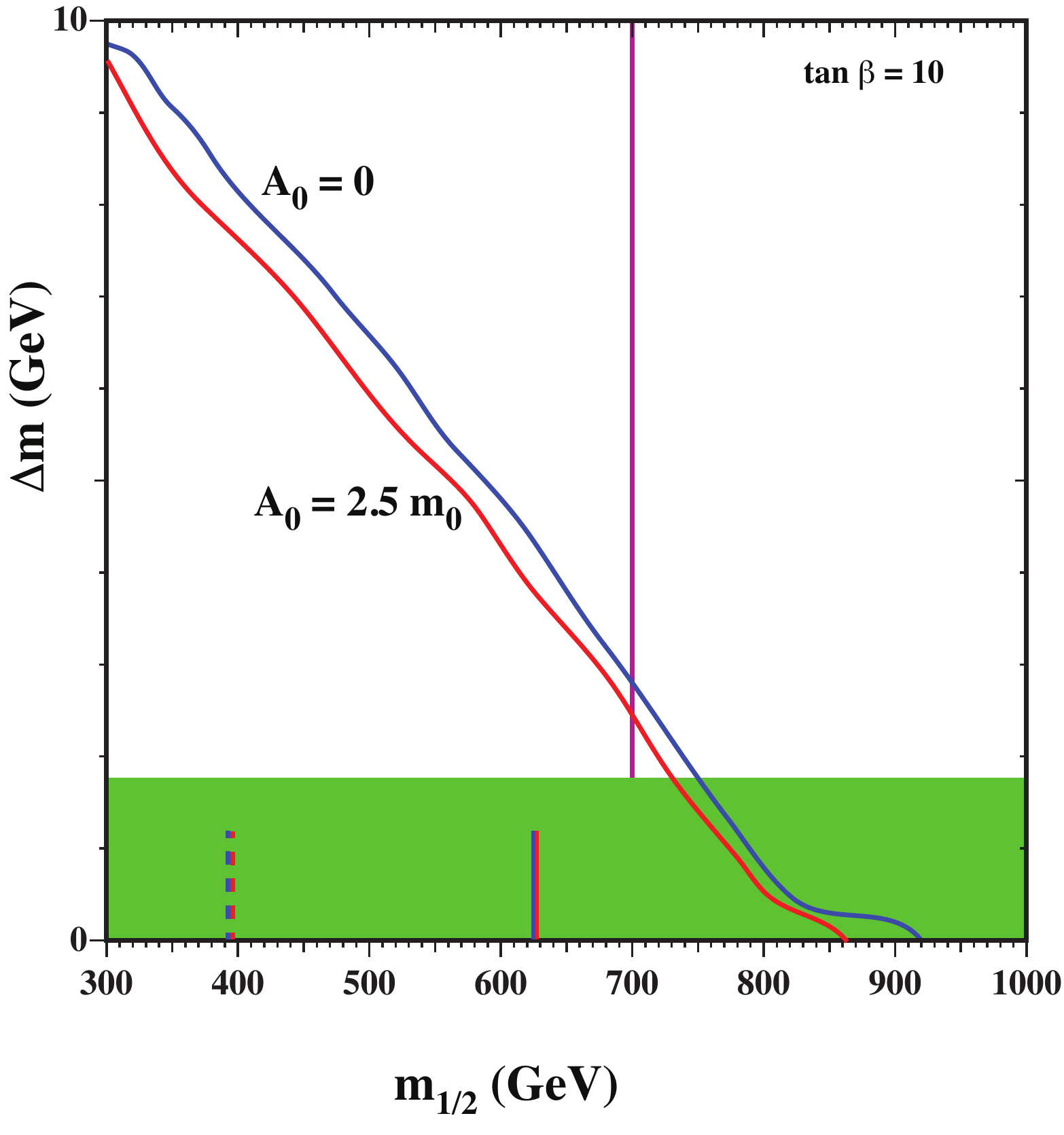}
   \includegraphics[height=2.65in]{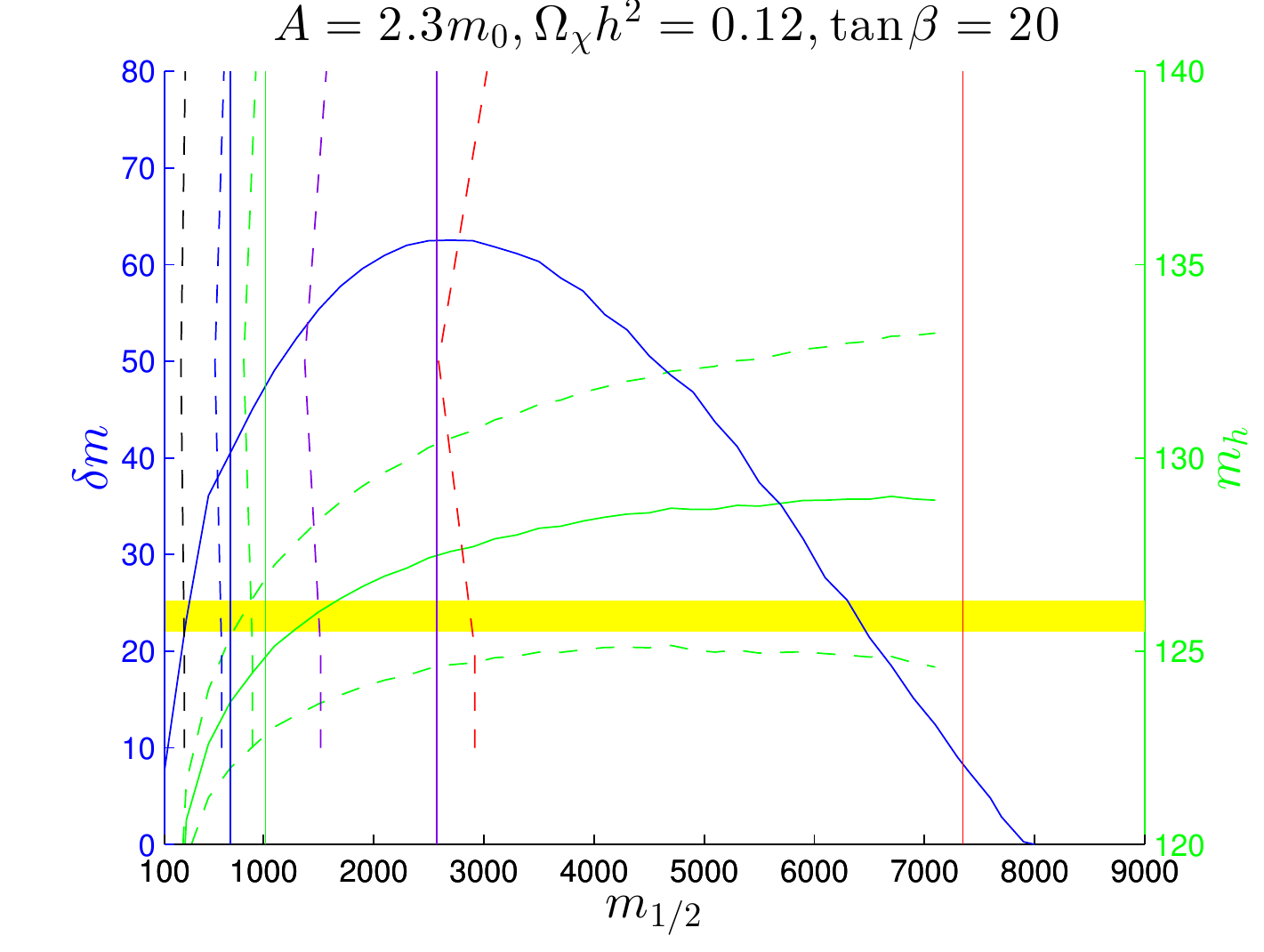}
  \caption{(left) The mass difference $\Delta m \equiv m_{\tilde \tau_1}-m_\chi$ 
as a function of $m_{1/2}$ along the CMSSM
coannihilation strips for $\tb = 10$ 
and for $A_0 = 0$ (blue line) and $2.5 \, m_0$ (red line).
The band with $m_{\tilde \tau_1}-m_\chi < m_\tau$ is shaded green.
The lower limit on $m_{1/2}$ from the 8-TeV ATLAS 5/fb $\ETslash$ search at the LHC~\protect\cite{ATLAS8}
is represented by the maroon line, and the lower limits from searches for the direct and total
production of metastable charged particles~\protect\cite{ATLASmcp} are shown as dashed and solid
lines, respectively, inside the green bands (see \protect\cite{celmov} for details).
(right) The solid blue curve is the profile in the $(m_{1/2}, \delta m \equiv m_{\tilde t_1} - m_\chi)$ plane
of the stop coannihilation strips for $A_0/m_0 = 2.3$ and $\tan \beta = 20$. 
The near-vertical black, blue, green, purple and red lines in each panel are
particle exclusion reaches for particle searches with LHC at 8~TeV,
300 and 3000/fb with LHC at 14~TeV, 3000/fb with HE-LHC at 33~TeV
and 3000/fb with FCC-hh at 100~TeV, respectively.
The solid lines are for generic $\ETslash$ searches, and the dashed lines are for dedicated
stop searches. The solid (dashed) near-horizontal green lines are central values (probable ranges)
of $m_h$, and the yellow band represents the experimental value of $m_h$ \cite{lhch}.
 }
  \label{strips}
\end{figure}

The extent of the stau coannihilation strip is shown in the left panel of Fig. \ref{strips}. 
There the stau-neutralino mass difference is plotted as a function of $m_{1/2}$ \cite{celmov}.
For this choice of $\tan \beta = 10$, the strip extends only out to $m_{1/2} \simeq 900$ GeV
and is all but excluded by the LHC (which excludes $m_{1/2} \lesssim 840$ GeV for relatively low $m_0$
\cite{ATLAS20}.
At $\tan \beta = 40$, the strip extends to slightly larger $m_{1/2} = 1300$ GeV when $A_0 = 2.5 m_0$.

When $A_0$ is large, one of the stop masses is driven small and the possibility for
stop coannihilation is realized \cite{stop-co,eoz,interplay,raza}. The stop coannihilation strip is also seen in the 
$(m_0,m_{1/2})$ plane in the left panel of Fig. \ref{mom12}. The stop strip corresponds to the thin
blue line which tracks the dark red wedge in the lower right of the panel. This strip extends
past the $m_0 = 10$ TeV extent of the figure. Unlike the stau strip, it is unlikely that the entire
strip will be fully probed as it is seen to extend beyond the reach of a future 33 TeV LHC upgrade.
The full extent of the stop strip is seen in the right panel of Fig. \ref{strips} which shows the stop-neutralino
mass difference as a function of $m_{1/2}$ \cite{eoz,interplay}. The stop strip extends to 
$\approx  8$ TeV in $m_{1/2}$ and $> 20$ TeV in $m_0$. 

\begin{figure}
  \includegraphics[height=.35\textwidth]{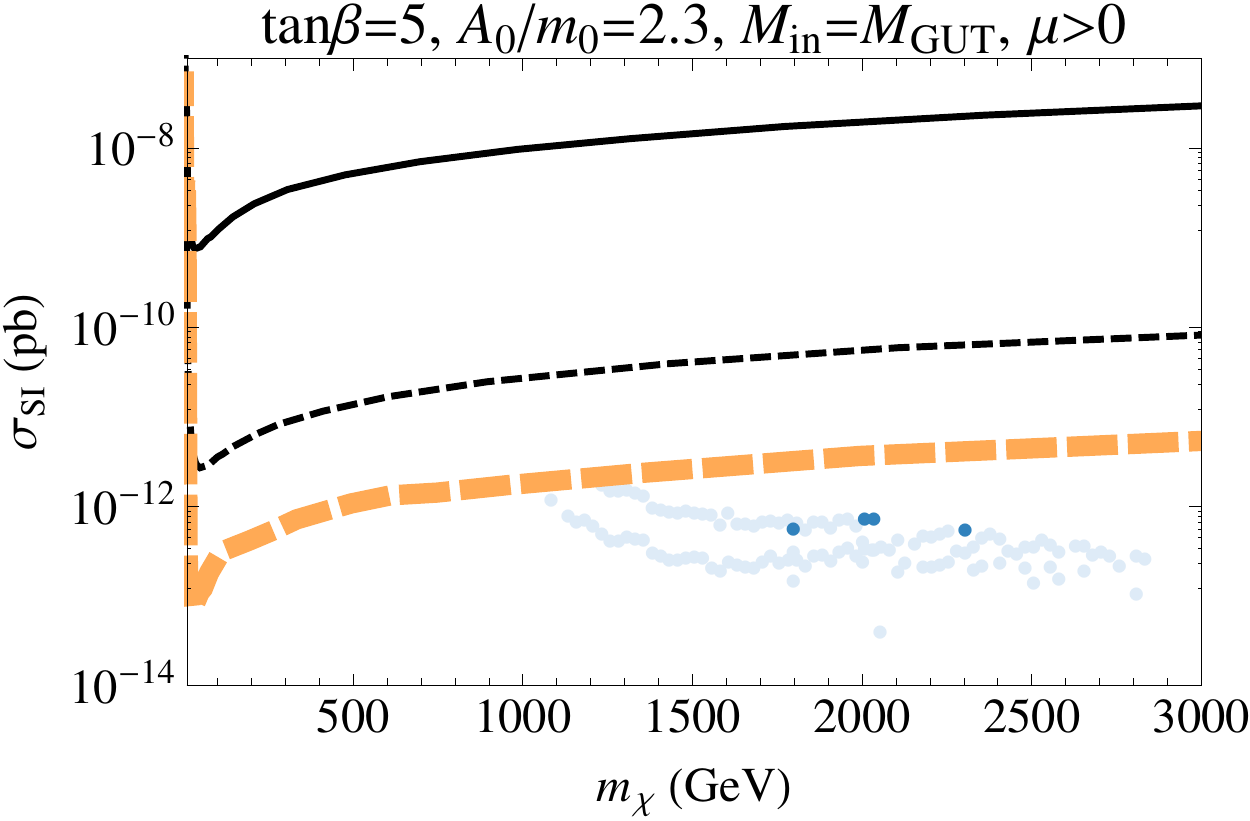}
   \includegraphics[height=.35\textwidth]{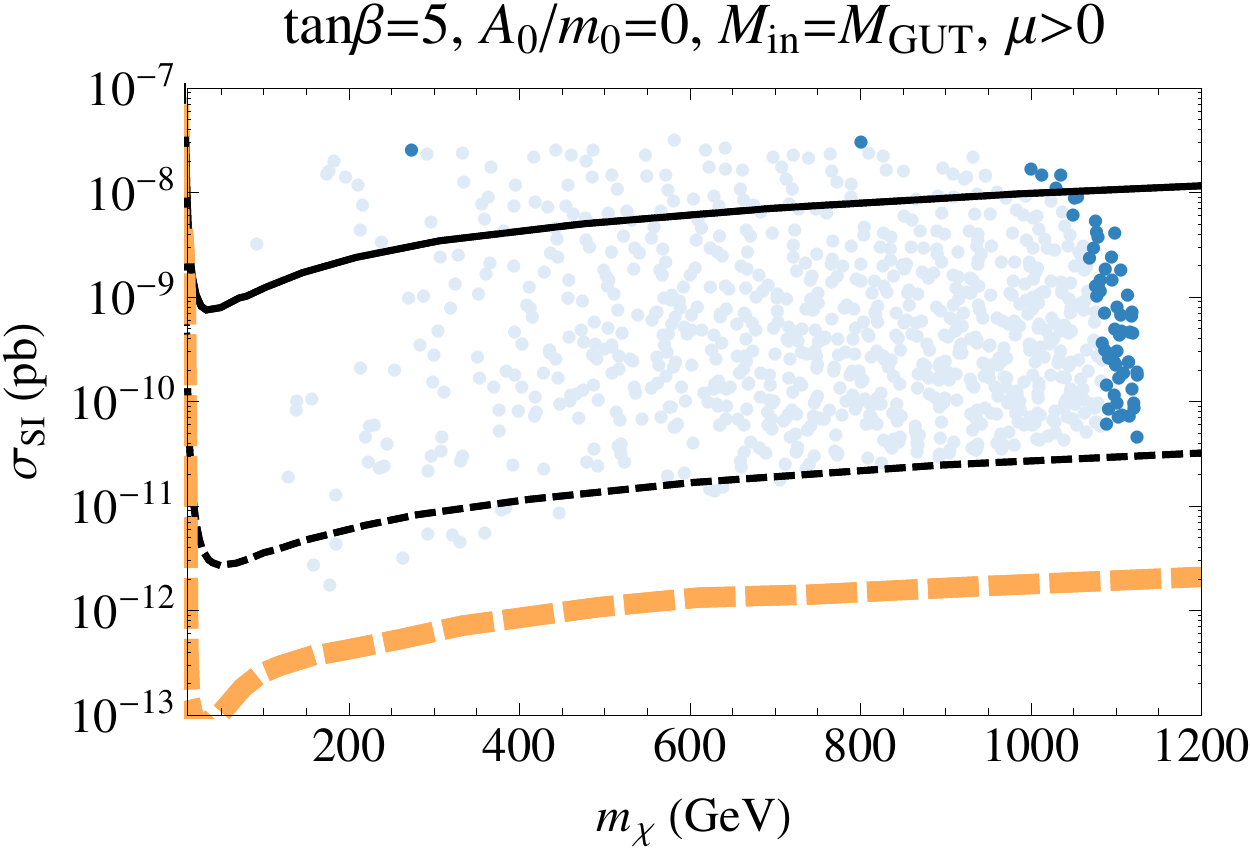}
  \caption{The spin-independent elastic scattering cross section in the
CMSSM as a function of the neutralino mass for $\mu > 0$, with
$\tan \beta = 5$ and $A_0 = 2.3\, m_0$ (left) and $A_0 = 0 $ (right).
The panels show points where the relic density is within 3$\sigma$ of
the central Planck value colored darker blue, and those where the relic
density is below the Planck value as lighter blue points. 
The black solid curve is the current LUX bound.
 The black dashed curve is the projected LZ sensitivity and
the dashed orange curve is the neutrino background level.
 }
  \label{ssi}
\end{figure}

Another difficulty for the stop strip seen in Fig. \ref{ssi} is its detectability in
direct detection experiments.  
The spin-independent cross section,
$\sigma_{\rm SI}$, as a
function of the neutralino mass  is shown in
the left panel of  Fig. \ref{ssi} for a similar case though
here $\tan \beta = 5$.
The points in the panel represent results of a scan of the parameter space.
Darker points fall within 3$\sigma$ of the dark matter relic density that fits best the Planck data \cite{Planck15}.
Lighter points have smaller relic densities and should not be excluded. However,
whenever the relic density is below the central value determined by Planck, we
scale the cross section downwards by the ratio of the calculated density
to the Planck density. The solid curve in Fig.~\ref{ssi} corresponds to
the current LUX limit \cite{lux}. The thin black dashed curve corresponds to the projected LZ  sensitivity \cite{LZ,cushman}. The thick orange dashed line corresponds
to the irreducible neutrino background \cite{nuback,cushman}.
In this case, the LSP is almost pure bino and squarks are quite heavy and as a result
the spin-independent cross section is quite small and generally falls below the neutrino background.

The focus point strip \cite{fp} occurs when $m_0$ is large and $A_0$ is relatively small.
In this case, the minimization of the Higgs potential yields a relatively small value for 
$\mu$ so that the LSP becomes Higgsino-like. An example of the focus point strip is seen
by the relatively thick blue strip in the right panel of Fig. \ref{mom12}.  To the right of the strip,
there is no solution to the Higgs minimization equations.
In contrast to the stop coannihilation strip, the focus point strip is clearly within
reach of direct detection experiments as seen in the right panel of Fig. \ref{ssi}. 
Almost all points sampled fall above the future reach of LZ \cite{LZ,cushman}, though
it should be noted that the constraint $m_h < 128$ GeV was imposed allowing one
to set a lower bound on the elastic cross section. 
The set of darkly shaded points with good relic density
are found mostly at $m_\chi \simeq 1100$ GeV due to the fact that these points are
mainly Higgsino LSPs \cite{osi}. 

\begin{figure}
  \includegraphics[height=.45\textwidth]{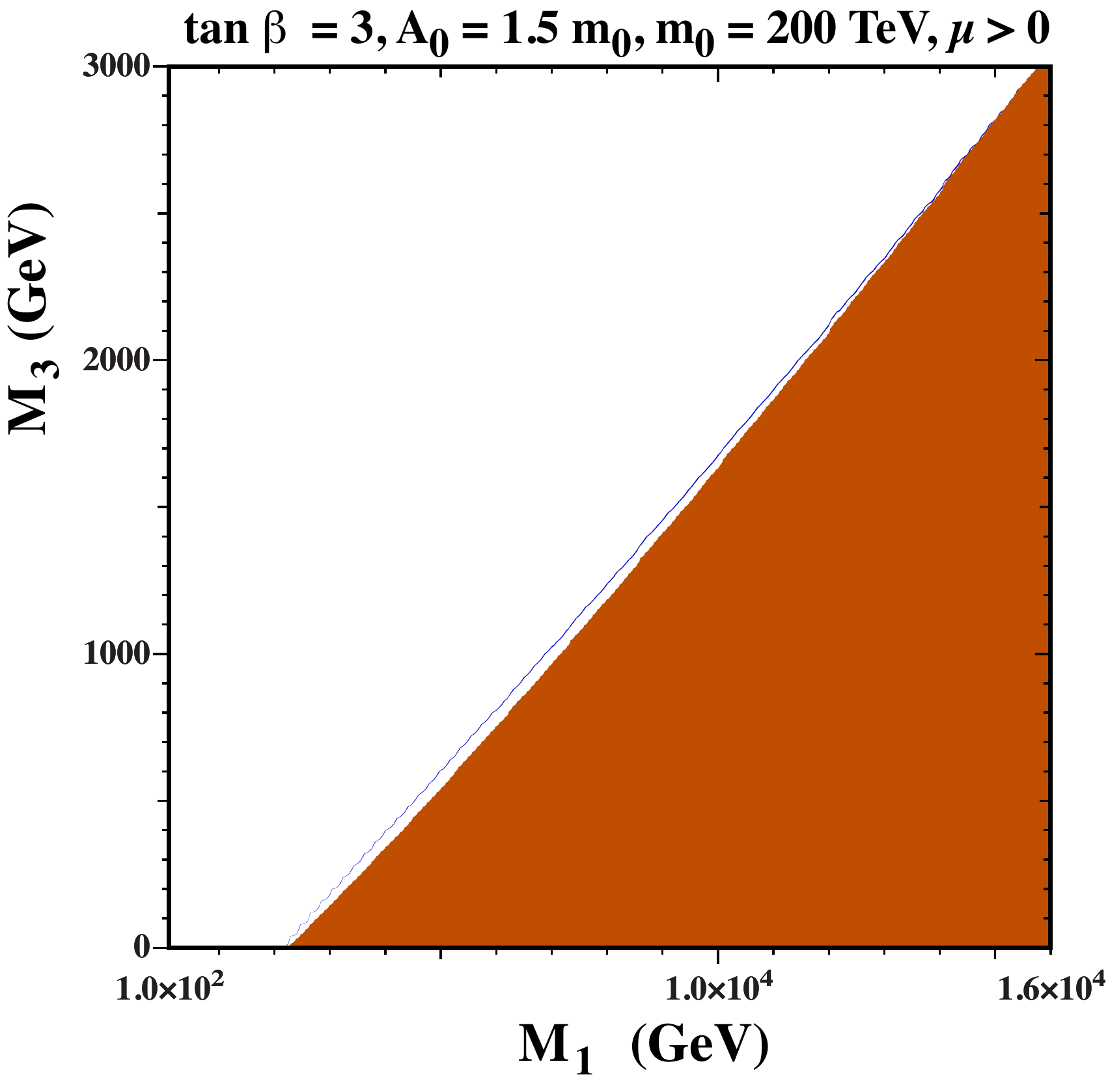}
   \includegraphics[height=.45\textwidth]{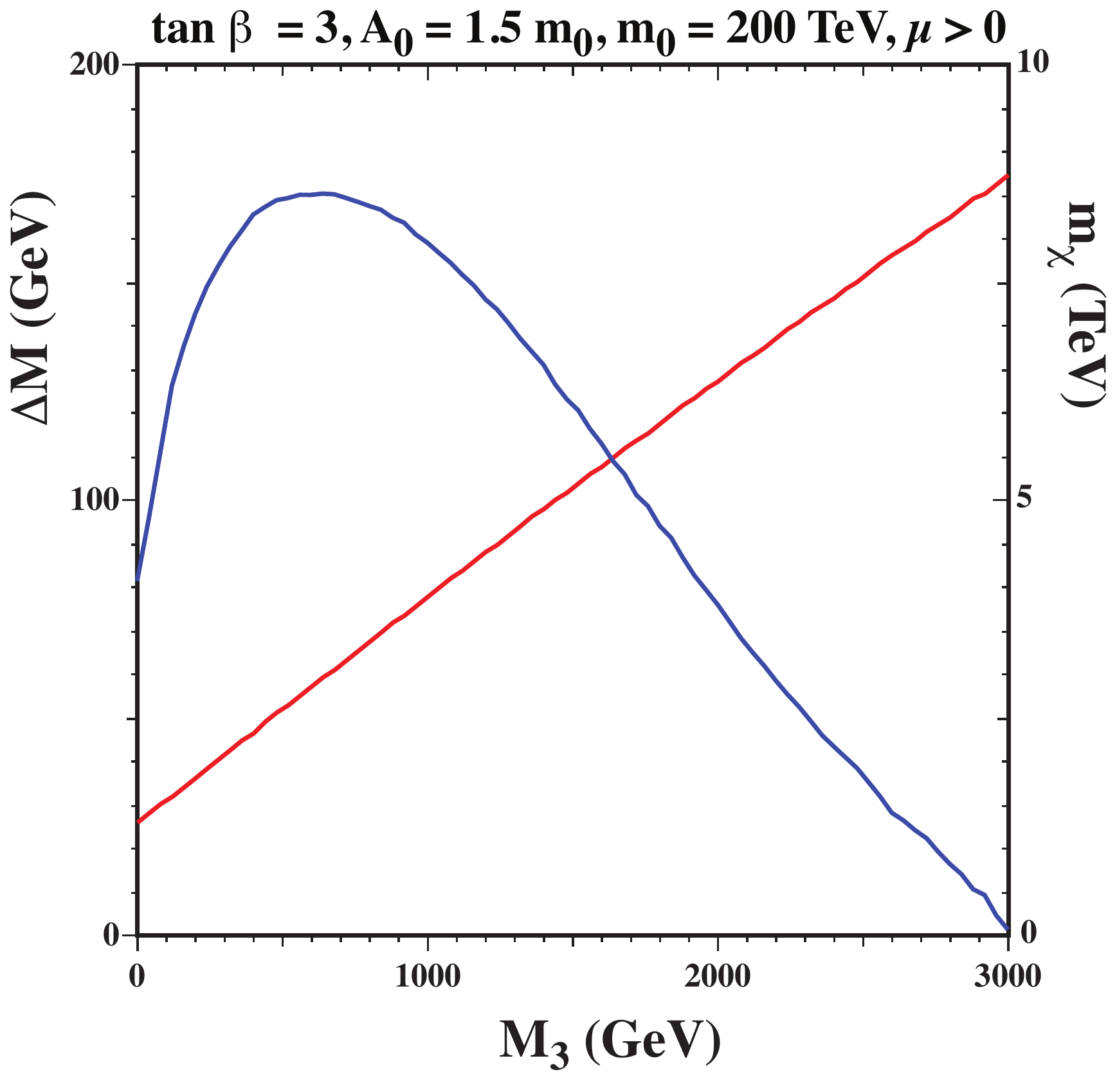}
  \caption{The $(M_1, M_3)$ plane (left) for $m_0 = 200$~TeV, $A_0/m_0 = 1.5$ and $\tan \beta = 3$.
The dark blue strip in the left panel shows where the relic
LSP density $\Omega_\chi h^2$ falls within the $\pm 3$-$\sigma$ range allowed by Planck
and other data, and the lightest neutralino is no longer the LSP in the regions shaded brick-red.
The right panel shows the gluino-neutralino mass difference (left axis, blue line) and the neutralino mass (right
axis, red line) as functions of $M_3$.
 }
  \label{m1m3}
\end{figure}

In the CMSSM, with non-universal gaugino masses, it is possible that gluino coannihilations
control the relic density \cite{glu,raza,ELO,EELO}. For example, by allowing the input gluino mass
to differ from the bino and wino masses at the GUT scale ($M_1 = M_2 \ne M_3)$,
the gluino may be the next lightest superpartner and opens the possibility for gluino coannihilations.
In the left panel of Fig. \ref{m1m3}, we show an example of a ($M_1,M_3$) plane for fixed 
$\tan \beta = 3$, $A_0/m_0 = 1.5$, and $m_0 = 200$ TeV \cite{EELO}. The gluino-bino coannihilation strip
is seen as the thin blue strip following the gluino LSP region shown in dark red and 
extends up to $\sim 3$ TeV. 
The panel on the right shows the gluino-neutralino mass difference $\Delta M$ (blue line) which peaks
at approximately 170 GeV, and is consistent with the results of \cite{ELO} for intermediate
squark-to-gluino mass ratios.  Also shown is the neutralino mass as a function of $M_3$ (red line):
it rises to $m_\chi \sim 8$~TeV at the tip of the coannihilation strip. The Higgs mass (which is not shown)
varies very slowly across the ($M_1,M_3$) plane and takes the value $m_h = 125$ for this value of $\tan \beta$. At lower values of $m_0$, more possibilities arise, namely, at large $M_1$, 
the composition of the LSP changes from bino to Higgsino, and the possibility of gluino-Higgsino
coannihilations appear. At still larger $M_1$ (for fixed $M_3$, there is a also a focus point region \cite{EELO}.

At large $\tan \beta$, it is also common that there are regions of the parameter space where
the LSP mass is very close to $m_A/2$, where $m_A$ is the Higgs pseudoscalar mass.
In this case, there is a large contribution to the cross section from rapid s-channel
annihilations, producing two strips (known as the funnel region) on either side of the pole \cite{funnel}.
However, at large $\tan \beta$, there are strong constraints from $B_s \to \mu^+ \mu^-$ decay~\cite{bmm}.
This region will not be discussed further here. 

There are of course many other possibilities in CMSSM-like models.
These include models where the Higgs mass are non-universal: the NUHM1 ($m_1 = m_2 \ne m_0$)
\cite{nuhm1,eosknuhm,elos,eelnos}, and NUHM2 ($m_1 \ne m_2 \ne m_0$)
\cite{nuhm2,eosknuhm,elos,eelnos}; subGUT models where the input universality scale differs from
the GUT scale with $M_{in} < M_{GUT}$ \cite{subGUT,elos,eelnos}; superGUT models with 
$M_{in} > M_{GUT}$ \cite{superGUT}. The above mentioned models all have 1-2 additional
parameters relative to the CMSSM. However there are also viable models with
fewer than the 4 free CMSSM parameters: These include mSUGRA models, where
the condition that $B_0 = A_0 - m_0$ is applied and hence $\tan \beta$ is no longer free
\cite{bfs,vcmssm,elos,eelnos}; and pure gravity mediated models \cite{pgm} for which
gaugino masses are generated through anomalies \cite{anom} on top of the mSUGRA conditions \cite{pgm2}
and produces a spectrum reminiscent of split supersymmetry \cite{split}. 

\section{SO(10) Dark Matter}

The motivations for supersymmetry are well known. 
These include the possibility for gauge coupling unification at the GUT scale \cite{Ellis:1990zq},
the stability of the electroweak vacuum \cite{Ellis:2000ig}, radiative electroweak symmetry breaking
\cite{ewsb}; a stable dark matter candidate \cite{ehnos}, the stabilization of the gauge hierarchy
\cite{hier}. With the exception of the latter, non-supersymmetric SO(10) GUT \cite{so10} models
may contain equivalences of all of these desirable features. In models with an intermediate scale
between the electroweak scale and the GUT scale, gauge coupling unification becomes possible
\cite{so10gc} when the intermediate scale is determined by the unification conditions
given a field content below the GUT scale. As discussed in more detail below, in
SO(10) models where the intermediate scale is broken by a Higgs in a {\bf 126} representation,
a residual $Z_2$ discrete symmetry survives enabling the possibility of 
a stable dark matter candidate \cite{so10dm,moqz,mnoqz,noz}. Furthermore, in models with
gauge coupling unification and a stable dark matter candidate, it is also possible to 
stabilize the electroweak vacuum while at the same time radiatively break
the electroweak symmetry \cite{mnoz}. In addition, one has the additional benefit of the seesaw
mechanism for generating neutrino masses \cite{seesaw}. 

To construct an SO(10) dark matter model, we should first pick an intermediate scale 
gauge group and a representation for the Higgs field, $R_1$ which breaks SO(10).
The possibilities considered are listed in Table \ref{tab:intgauge}.

\begin{table}[ht!]
 \begin{center}
\caption{\it Candidates for the intermediate gauge group $G_{{int}}$.}
\label{tab:intgauge}
\vspace{5pt}
\begin{tabular}{ll}
\hline
\hline
$G_{{int}}$ & $R_1$ \\
\hline
${SU}(4)_C\otimes {SU}(2)_L \otimes {SU}(2)_R$& {\bf 210}\\
${SU}(4)_C\otimes {SU}(2)_L \otimes {SU}(2)_R\otimes {D}$& {\bf
     54}\\
${SU}(4)_C\otimes {SU}(2)_L \otimes {U}(1)_R$ & {\bf 45}\\
${SU(3)}_C\otimes {SU}(2)_L \otimes {SU}(2)_R
 \otimes {U}(1)_{B-L}$ &{\bf 45}\\
${SU(3)}_C\otimes {SU}(2)_L \otimes {SU}(2)_R
 \otimes {U}(1)_{B-L} \otimes D$ & {\bf 210}\\
${SU(3)}_C\otimes {SU}(2)_L \otimes {U}(1)_R 
 \otimes {U}(1)_{B-L}$ & {\bf 45}, {\bf 210}\\
\hline
${SU}(5) \otimes {U}(1)$ & {\bf 45}, {\bf 210}\\
${Flipped}~ {SU}(5) \otimes {U}(1)$ & {\bf 45}, {\bf 210}\\
\hline
\hline
\end{tabular}
 \end{center}
\end{table}

As noted above, we must employ a {\bf 126} to break the intermediate gauge group
down to the SM in order to preserve a $Z_2$ symmetry related to matter parity. 
The coupling of the {\bf 126} to SM matter fields embedded in a {\bf 16}
representation of SO(10) naturally gives rise to a majorana mass mass to the 
$\nu_R$ component of the {\bf 16} of order $\langle {\bf 126} \rangle \sim M_{int}$ which when
combined with the Dirac mass arising from the vev of the SM Higgs (now residing in a 10-plet of SO(10))
gives rise to the seesaw mechanism for light neutrino masses \cite{seesaw}. 

Next we must choose a representation for the dark matter candidate.
Possible choices are given in Table \ref{tab:dmcandidate}.
A fermionic DM candidate should be
parity even and belong to a ${\bf 10}$, ${\bf 45}$, ${\bf 54}$,
${\bf 120}$, ${\bf 126}$, ${\bf 210}$ or ${\bf 210}^{\prime}$
representation, while scalar DM is 
parity odd and belongs to a $\bf 16$ or $\bf 144$
representation. Following the branching rules given in
Ref.~\cite{Slansky:1981yr}, in Table~\ref{tab:dmcandidate}, we list
${SU(2)}_L\otimes{U(1)}_Y$ multiplets in various SO(10)
representations that contain an electrically neutral color singlet.
The table is classified by $B-L$ so one can
check the matter parity of the candidates easily; $B-L=0,~2$ candidates
are fermionic while $B-L=1$ candidates are scalar, labeled by an ``{\tt
F}'' or ``{\tt S}'' at the beginning of each row, respectively. The
subscript of the model names denotes the ${SU}(2)_L$
representation, while the superscript shows hypercharge. A hat is used for
$B-L = 2$ candidates.

\begin{table}[t]
 \begin{center}
\caption{\it List of ${SU(2)}_L\otimes{U(1)}_Y$ multiplets in
  SO(10) representations that contain an electric neutral color singlet. }
\label{tab:dmcandidate}
\vspace{5pt}
\begin{tabular}{l c c l l}
\hline
\hline
 Model & $B-L$ & ${SU(2)}_L$ & $Y\qquad$ &  SO(10) representations\\
\hline
\model{F}{1}{0} &\multirow{6}{*}{0}
 & {\bf 1} & $0$ &   {\bf 45}, {\bf 54}, {\bf 210} \\
\model{F}{2}{1/2} & & {\bf 2} & $1/2$ & {\bf 10}, {\bf 120}, {\bf 126}, ${\bf 210}^\prime$ \\
\model{F}{3}{0}& & {\bf 3} & $0$ &  {\bf 45}, {\bf 54}, {\bf 210} \\ 
\model{F}{3}{1}& & {\bf 3} & $1$ &   {\bf 54} \\ 
\model{F}{4}{1/2}& & {\bf 4} & $1/2$ &  ${\bf 210}^\prime$ \\
\model{F}{4}{3/2}& & {\bf 4} & $3/2$ &  ${\bf 210}^\prime$ \\
\hline
\model{S}{1}{0} & \multirow{4}{*}{1}
 & {\bf 1} & $0$ &   {\bf 16}, {\bf 144} \\
\model{S}{2}{1/2}& & {\bf 2} & $1/2$&    {\bf 16}, {\bf 144} \\
\model{S}{3}{0}& & {\bf 3} & $0$&  {\bf 144} \\
\model{S}{3}{1}& & {\bf 3} & $1$&  {\bf 144} \\ 
\hline
\model{\widehat{F}}{1}{0}& \multirow{3}{*}{2}
 & {\bf 1} & $0$ &   {\bf 126} \\
\model{\widehat{F}}{2}{1/2}& & {\bf 2} & $1/2$&   {\bf 210} \\
\model{\widehat{F}}{3}{1}& & {\bf 3} & $1$& {\bf 126} \\
\hline
\hline
\end{tabular}
 \end{center}
\end{table}

\begin{figure}
\begin{center}
  \includegraphics[height=.45\textwidth]{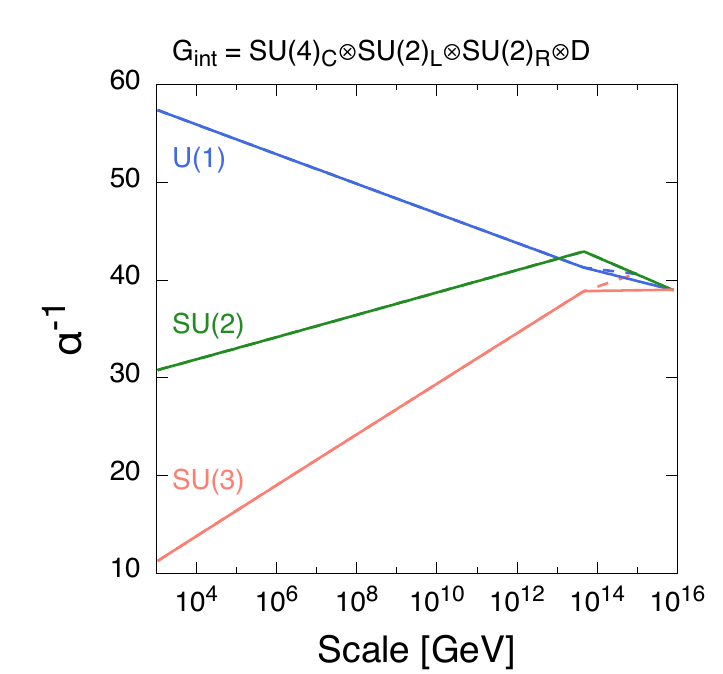}
  \end{center}
  \caption{Running of the gauge couplings for the fermionic singlet dark matter model 
  described in the text.
 }
  \label{running}
\end{figure}

Depending on the dark matter and Higgs representation chosen, renormalization 
group evolution of the gauge couplings can be used to determine,\
the GUT scale, the intermediate scale, and the value of the GUT gauge coupling.
One such example is a singlet fermion (\model{F}{1}{0}) originating in the ({\bf 15, 1, 1}) representation 
(in terms of ${SU}(4)\otimes {SU}(2) \otimes {SU}(2)$) included in the {\bf 45} of SO(10). 
The evolution of the gauge couplings in this model is shown in Figure \ref{running}. 
In this model \cite{mnoqz}, $R_1 = $ {\bf 54},
and we have $\log (M_{int}) = 13.66, \log (M_{GUT}) = 15.87$, and $g_{GUT} = 0.567$. 
For further details concerning this model, see \cite{mnoqz}.

Given the large number of possible intermediate gauge groups and the large number of possible
dark matter representations (scalar or fermionic), one may think that there are a vast
number of dark matter models in the SO(10) framework. However,
once we demand gauge coupling unification, there are in fact only a handful of models
which permit gauge coupling unification and satisfy constraints from the lifetime of the proton.
We also require that the solution give $M_{int} < M_{GUT}$
and that we can split the SO(10) multiplets in such a way to leave
only the DM candidate (and perhaps weak partners) with weak scale masses. 
The resulting acceptable models
for scalar dark matter candidates is shown in Table \ref{scalar}. For more information on these models
see \cite{noz}. The singlet models (SA) have a phenomenology similar to that of so-called
Higgs portal models~\cite{Burgess:2000yq}. To roughly estimate the favored
mass region for a scalar singlet, consider the quartic
interaction between the singlet DM $\phi$ and the SM Higgs field:
$\lambda_{H\phi} \phi^2 |H|^2/2$. Through this coupling, the singlet DM
particles annihilate into a pair of the SM Higgs bosons and weak gauge bosons. The
annihilation cross section  is
\begin{equation}
\sigma_{{ann}}v_{{rel}} \simeq
\frac{\lambda^2_{H\phi}}{16 \pi m_{{DM}}^2}~,
\end{equation} 
assuming that the DM mass $m_{{DM}}$ is much larger than the
SM Higgs mass $m_h$ and neglecting terms proportional to 
$v^2$. The DM relic abundance is, on the other hand,
related to the annihilation cross section by
\begin{equation}
\Omega_{{DM}} h^2 \simeq  
\frac{3 \times 10^{-27}~{\rm cm}^3 ~{\rm s}^{-1}}{\langle
 \sigma_{{ann}} v_{{rel}} \rangle} ~.
\label{eq:DMab}
\end{equation}
To account for the observed DM density $\Omega_{{DM}} h^2 = 0.12$~\cite{Planck15},
the DM mass should be  $m_{{DM}}\lesssim 10~{\rm TeV}$ for
$\lambda_{H\phi}\lesssim 1$. This gives us a rough upper bound for the DM mass.   
The other scalar DM candidates are ${SU(2)}_L\otimes {U}(1)_Y$
multiplets, which can interact with SM particles through gauge
interactions in addition to the quartic coupling mentioned above. In particular,
\model{S}{2}{1/2} is known as the Inert Higgs Doublet Model and has been widely
studied in the literature
\cite{Deshpande:1977rw, Arhrib:2013ela}.

\begin{table}[ht]
\centering
\caption{
\it One-loop result for $M_{GUT}$, $M_{int}$,
$\alpha_{GUT}$, and proton lifetimes
 for scalar dark matter models. The DM mass is set to be
 $m_{DM}= 1~{TeV}$. The mass scales are given in GeV and the
 proton lifetimes are in units of years. These models evade the
 proton decay bound, $\tau(p\rightarrow e^+ \pi^0)>1.4\times
10^{34}~{yrs}$\cite{Shiozawa, Babu:2013jba}.}
\label{scalar} 
\vspace{5pt}
\scalebox{0.93}{
  \begin{tabular}{llllll}
\hline
\hline
 Model  & $R_{\rm DM}$  & $\log_{10} M_{\rm GUT}$  & $\log_{10}M_{\rm int}$ &
   $\alpha_{\rm GUT}$  & $\log_{10}\tau_p(p\rightarrow e^+ \pi^0)$ \\ 
    \hline  
 \hline 
    \multicolumn{5}{c}{$G_{\rm int}={SU}(4)_C\otimes
   {SU}(2)_L\otimes {SU}(2)_R$}\\
 \hline
 \DM{SA}{422} (\model{S}{1}{0} ) & ${\bf 4},{\bf 1},{\bf 2}$ &
  $16.33$ & $11.08$ & $0.0218$~~~~ & $ 36.8 \pm 1.2$ \\ 
 \DM{SB}{422}  (\model{S}{2}{1/2} )& ${\bf 4},{\bf 2},{\bf 1}$ &
  $15.62$ & $12.38$ & $0.0228$ & $ 34.0 \pm 1.2$ \\ 
     \bhline{1.5pt}
    \multicolumn{5}{c}{$G_{\rm int}={SU}(3)_C\otimes
   {SU}(2)_L\otimes {SU}(2)_R \otimes {U}(1)_{B-L}$}\\
    \hline
\DM{SA}{3221}  (\model{S}{1}{0} ) & ${\bf 1},{\bf 1},{\bf 2},{ 1}$ &
$16.66$ & $ 8.54$ & $0.0217$ & $ 38.1 \pm 1.2$ \\ 
\DM{SB}{3221}  (\model{S}{2}{1/2} ) & ${\bf 1},{\bf 2},{\bf 1},{ -1}$ &
$16.17$ & $ 9.80$ & $0.0223$ & $ 36.2 \pm 1.2$ \\ 
\DM{SC}{3221}  (\model{S}{2}{1/2} )  & ${\bf 1},{\bf 2},{\bf 3},{ -1}$ & 
$15.62$ & $ 9.14$ & $0.0230$ & $ 34.0 \pm 1.2$ \\  
\bhline{1.5pt}
     \multicolumn{5}{c}{$G_{\rm int}={SU}(3)_C\otimes
   {SU}(2)_L\otimes {SU}(2)_R \otimes {U}(1)_{B-L}\otimes D$}\\
    \hline
\DM{SA}{3221D} (\model{S}{1}{0} ) & ${\bf 1},{\bf 1},{\bf 2},{ 1}$ & 
$15.58$ & $10.08$ & $0.0231$ & $ 33.8 \pm 1.2$ \\ 
\DM{SB}{3221D} (\model{S}{2}{1/2} ) & ${\bf 1},{\bf 2},{\bf 1},{ -1}$ & 
$15.40$ & $10.44$ & $0.0233$ & $ 33.1 \pm 1.2$ \\   
 \hline 
 \hline
  \end{tabular}
}
\end{table}

An example of a fermionic singlet was discussed above, and its relic density is mainly
determined by the reheat temperature after inflation (see \cite{moqz,mnoqz} for more details).
Non-singlet fermions behaving as well studied wimps are also possible. From Table \ref{tab:dmcandidate},
we see that the only candidates without hypercharge is the weak (wino-like) triplet \model{F}{3}{0}.
One example of a triplet candidate is given in Table \ref{triplet}. As one can see,
this state is a singlet under both SU(4)$_C$ and SU(2)$_R$ and originates in a {\bf 45} of SO(10).
While the intermediate scale is relatively low, the GUT scale is quite high 
and hence the proton lifetime is unobservably long.

\begin{table}[ht!]
\centering
\caption{\it The one-loop results for $M_{GUT}$, $M_{int}$,
$\alpha_{GUT}$, and proton lifetimes for a real triplet fermionic DM
 models. Here the DM mass is set to be 1~TeV. The mass scales and proton
 decay lifetime are in units of GeV and years, respectively. } 
\label{triplet}
\vspace{5pt}
  \begin{tabular}{llcccc}
    \hline
    \hline
Model & $R_{\rm DM}$ &  $\mathrm{log}_{10} M_{\rm int} $ & 
$\mathrm{log}_{10} M_{\rm GUT} $ & $\alpha_{\rm GUT}$ &  $\log_{10}\tau_p(p\rightarrow e^+ \pi^0)$
 \\
 \hline 
\hline
 \multicolumn{6}{c}{$G_{int}={\rm SU(4)}_C\otimes{\rm SU(2)}_L\otimes{\rm
SU(2)}_R$}\\
\hline
\model{{F}}{3}{0} & $({\bf 1},{\bf 3},{\bf 1})$ & $ 6.54$ & $17.17$ & $0.0252$ & $39.8\pm 1.2$\\ 
 \hline
 \hline
  \end{tabular}
\end{table}

It is also possible that a fermionic dark matter candidate carries hypercharge and in this case, it may
be either a weak doublet (Higgsino-like) or triplet. 
Some examples are shown in Table \ref{hyper}. In this case, we must introduce another
representation (at the intermediate scale) to mix with $R_{\rm DM}$ in order to induce
some splitting in the DM multiplet to evade current DM detection experimental results. 
These are denoted $R^\prime_{\rm DM}$ in the table. 

\begin{table}[ht!]
\centering
\caption{\it \label{hyper}
Possible hypercharged fermionic DM models that are not yet excluded by
 current proton decay experiments. The quantum numbers are labeled in
 the same order as $G_{int}$. The subscripts D and W refer
 to Dirac and Weyl respectively.  The numerical results are calculated
 for DM mass of 1~TeV. The mass scales and proton decay lifetime are in
 unit of GeV and years, respectively. }
  \vspace{5pt}
{
  \begin{tabular}{cllllll }
    \hline
    \hline
 Model & $R_{\rm DM}$ & $R^\prime_{\rm DM}$ & $\mathrm{log}_{10} M_{\rm int} $ & 
$\mathrm{log}_{10} M_{\rm GUT} $ & $\alpha_{\rm GUT}$ &
   $\log_{10}\tau_p $
 \\
\hline
 \hline
 \multicolumn{7}{c}{$G_{int}={\rm SU(4)}_C \otimes{\rm SU(2)}_L\otimes{\rm U(1)}_R$}
 \\
 \hline 
 \DM{FA}{421} (\model{{F}}{2}{1/2}) & $({\bf 1},{\bf 2},1/2)_D$ 
 & $({\bf 15},{\bf 1},0)_W$ 
 & $3.48$  & $17.54$ & $0.0320$ 
 & $40.9\pm 1.2$ \\
 \bhline{1.5pt}
  \multicolumn{7}{c}{$G_{int}={\rm SU(4)}_C\otimes{\rm
   SU(2)}_L\otimes{\rm SU(2)}_R$} 
 \\
 \hline 
 \DM{FA}{422} (\model{{F}}{2}{1/2})& $({\bf 1},{\bf 2},{\bf 2})_W$ 
 & $({\bf 1},{\bf 3},{\bf 1})_W$ 
  & $9.00$  & $15.68$ & $0.0258$ 
 & $34.0\pm 1.2$ \\
 \hline 
 \DM{FB}{422} (\model{{F}}{2}{1/2})& $({\bf 1},{\bf 2},{\bf 2})_W$ 
 & $({\bf 1},{\bf 3},{\bf 1})_W$ 
 & $5.84$  & $17.01$ & $0.0587$ 
 & $38.0\pm 1.2$ \\
 \hline
 \hline
  \end{tabular}
}
\end{table}

\section{Summary}
It is becoming apparent that recent LHC searches for supersymmetry have pushed
CMSSM into corners of the parameter space which rely on the near degeneracy between the LSP
and the next lightest superpartner, thus allowing coannihilations to reign in the relic density.
While the stau coannihilation strip is nearly ruled out by LHC searches, possibilities remain
for the stop strip and if there are non-universal gaugino masses, gluino coannihilation.
It is also possible that $m_0$ is large near the focus point strip so that the LSP is mostly Higgsino-like.
Though not discussed here, there are several variants of the CMSSM which still permit
neutralino dark matter. These include models with non-universal Higgs scalar masses (NUHM),
models where the input universality scale is below the GUT scale (subGUT models), 
or pure gravity mediated models with either wino or Higgsino dark matter.

While supersymmetry has many motivations beyond dark matter, 
with the exception of the hierarchy problem, almost of these motivating factors
can be resolved in non-supersymmetric version of SO(10) grand unifications.
Several such examples were outlined above.
The real challenge lies in the detection of dark matter and our ability to 
discriminate between the various models.

\end{document}